\documentstyle[12pt]{article}

\newcommand{\beq}{\begin{equation}}
\newcommand{\eeq}{\end{equation}}

\title{
Comments to  {\sl On the Accuracy of Lamb Shift Measurements in Hydrogen} 
(Physica Scripta,  55 (1997) 33--40)
by V. G. Pal'chikov, Yu. L. Sokolov, and V. P. Yakovlev
}

\author{
Savely G. Karshenboim\thanks{
E-mail: sgk@onti.vniim.spb.su;
karshenboim@phim.niif.spb.su}
\bigskip
{}\\
 D. I.  Mendeleev Institute for Metrology,\\
198005, St.~Petersburg, Russia
}

\date{}

\begin{document}

\large

\maketitle

The major part of calculation in this work (\cite{Pal97}) is focused on the decay width evaluation of the
2p-state. They are contained in the appendix of the article.
The current work seems to be interesting in view of existing discrepancy
for the $2p$-state lifetime calculation 
between the former work \cite{Pal83} of the present authors and the results of Ref.
\cite{JETP94}.

It should be pointed out in this respect that the result of the
paper under consideration are obtained within the framework of the
nonrelativistic approach, whereas there exists a fully relativistic
expression being cited by the authors (see eqs.(9,10)). The relativistic formula
involves the terms containing derivatives 

$$
\Gamma^{(2)}(a)\,\frac{\partial}{\partial E}
\langle a\vert\Sigma^{(2)}(E)\vert a\rangle\bigg\vert_{E=E_a}, 
$$

\noindent
which are absent in its
nonrelativistic analogue.

Furthermore, the calculationsi, which are carried out  in
the paper, are performed without specifying the gauge. This circumstance is
rather important, since the mass operator is calculated off the mass shell.
Thus,  some diagrams in the Feynmann gauge contribute to the order of $\alpha$ in units of
the non-relativistic contributuion. It is
therefore obvious that the use of a fully  relativistic treatment  will be
strongly required so as to obtain corrections of the order of $\alpha(Z\alpha)^2$.

One should also mention that the nonrelativistic formalism is valid solely  in
two particular cases. Namely, in the vacuum polarization calculation, as  well
as  in obtaining the logarithmic part of the self-energy contribution in the
Yennie gauge where the self-energy is described by means of the local
potential (see Ref. \cite{JETP94} for details).

In the course of discussion dealing with the above-mentioned contributions,
the inaccuracy of the answer is obvious. The authors' result reduces
then to the statement that the nonrelativistic formula to define the width
of the level involves only corrections to the photon's frequency, rather
than to the dipole matrix element, provided that the vacuum polarization 
is described by only the local potential, whereas the operator of the vertex-type is absent.

The aforementioned result is closely connected with  innacuracy while
proceeding from eq.(16) to eq.(17). In particular, the work contains some misprints with
one being the absence of the subscripts in eq.(16). After
correcting these relations accordingly, eq. (16) should then take the form

$$
M^{RAD}=-\frac{e}{(2\omega)^{1/2}}\langle 2p_{1/2}\vert\bigg\{
\sum_{n\neq 2p_{1/2}}{
\frac{\Sigma^{(2)}(E_{2p_{1/2}})\vert n\rangle\langle n\vert
\mbox{$\boldmath{\varepsilon\cdot \alpha}$} \, 
exp(iqx)}{E_{2p}-E_{n}}
}$$

$$
+
\sum_{n\neq 1s_{1/2}}{
\frac{\mbox{$\boldmath{\varepsilon\cdot \alpha}$} \,
 exp(iqx)\vert n\rangle
\langle n\vert\Sigma^{(2)}(E_{1s})}{E_{1s}-E_{n}}
}
\bigg\}\vert 1s_{1/2}\rangle
.$$

\noindent
That is the relativistic expession, but one can find the nonrelativistic analogue.

The denominator in the first term is equal to $E_{2p}-E_{n}$ and that is due to 
calculation of the correction to the $2p$-state wave function.
However, 
using of the commutator of eq.(13) in the nominator matrix element between $1s$ and 
$n$-states leads
to energy difference $E_{1s}-E_{n}$ (cf. intermediate step from eq. (12) to eq. (14)). This factor 

$$\frac{E_{1s}-E_{n}}{E_{2p}-E_{n}}=1+\frac{E_{1s}-E_{2p}}{E_{2p}-E_{n}}
$$

\noindent can
never be equal to one and the result of eq.(18) can not be obtained.  
Some confusuion with the indexes has lead, therefore, to the mutual
cancellation of completely different quantities. The result of the paper can not be also obtained in 
any approximation, as far as any energy values ($E_{1s},E_{2p}$ and $E_{n}$) are of the same order of magnitude. 

The term of "1" in the equation above corresponds to the first term of eq.(17) 
and leads to the shift of the energy of the $2p$-state. The second term, which 
has been lost in the paper \cite{Pal97}, associated with the correction to the
 $2p$-state wave function and leads to correction to the dipole matrix element,
 which is included in result of Ref. \cite{JETP94} and exluded in Ref.  \cite{Pal97}. 
Conversely, one
would get the result, analogous to that of Ref. \cite{JETP94} to define the vacuum
polarization contribution, by recovering the indexes in the aforementioned
expression. Furthermore, it should be noted that the analytic 
 results obtained in Ref. \cite{IK96} can readily be used for the vacuum
polarization calculation.

It should be emphasised, however, that our criticism concerns only the
value of the lifetime employed, but not the main idea of the given experiment.

As for the calculation of the constant term, then it should be mentioned
that\\
(i) its correct definition can be achieved solely within the framework
    of the fully relativistic approach;\\
(ii) the appropriate correction to the final results of the work
     may depend on the process considered;\\
(iii) correction of the same order arise also when one would allow for
     the line shape. The latter was decribed in Ref. \cite{Pal83} 
by using the simple
   Lorenz contour.\\

By summarizing the given arguments, one can state that the result 
of Refs. \cite{Pal83,Pal97}
constitutes the superfluous accuracy (which originates from the
nonrelativistic treatment), as well as contains an apparent mistake.
By correcting the latter, one ends up, as it have already been pointed out, with
the result of Refs. \cite{JETP94,IK96} being argued by the authors.


\begin{thebibliography}{9}
\frenchspacing

\bibitem{Pal97} V. G. Pal'chikov, Yu. L. Sokolov, V. P. Yakovlev,
Physica Scripta, {\bf 55} (1997) 33.

\bibitem{Pal83} V. G. Pal'chikov, Yu. L. Sokolov, V. P. Yakovlev,
Pis'ma ZhETF {\bf 38} (1983) 347 /in Russian/; JETP Letters {\bf 38}
(1983) 418; Metrologia {\bf 21}, 99 (1985).

\bibitem{JETP94} S. G. Karshenboim, ZhETF {\bf 106} (1994) 414 /in
Russian/; JETP {\bf 79} (1994) 230; Yad. Fiz. {\bf 58} (1995)
901 /in Russian/; Phys. At. Nucl. {\bf 58} (1995) 835;
ZhETF {\bf 107} (1995) 1061 /in
Russian/; JETP {\bf 80} (1995) 593.

\bibitem{IK96} V.~G.~Ivanov and S.~G.~Karshenboim, Phys. Lett. {\bf A210}
(1996) 313; ZhETF {\bf 109} (1996) 1219 /in Russian/; JETP {\bf 82} (1996)
656.

\end{thebibliography}
\end{document}